\documentclass[aps,prl,twocolumn,includegraphics,preprintnumbers,superscriptaddress]{revtex4}

\usepackage{slashed}
\usepackage{epsfig,latexsym,cancel,amssymb,amsmath,verbatim,mathrsfs}
\usepackage{color}
\usepackage{graphicx}
\usepackage{amssymb} 
\usepackage{amsmath}
\usepackage{bbold}
\usepackage{slashed}
\usepackage{hyperref}
\usepackage{verbatim}
\usepackage{multirow}
\usepackage{chemarrow}
\usepackage{float}
\usepackage{stmaryrd}

\newcommand{\ET}{\mbox{$\not \hspace{-0.10cm} E_T$ }}

\newcommand{\beq}{\begin{equation}}
\newcommand{\eeq}{\end{equation}}
\newcommand{\bea}{\begin{eqnarray}}
\newcommand{\eea}{\end{eqnarray}}
\newcommand{\ba}{\begin{array}}
\newcommand{\ea}{\end{array}}
\newcommand{\bi}{\begin{itemize}}
\newcommand{\ei}{\end{itemize}}
\newcommand{\bn}{\begin{enumerate}}
\newcommand{\en}{\end{enumerate}}
\newcommand{\bc}{\begin{center}}
\newcommand{\ec}{\end{center}}
\renewcommand{\l}{\left}
\renewcommand{\r}{\right}

\newcommand{\eq}[1]{Eq.~(\ref{#1})}

\newcommand{\GeV}{\mathinner{\mathrm{GeV}}}
\newcommand{\TeV}{\mathinner{\mathrm{TeV}}}
\newcommand{\met}{\slashed{E}_T}

\begin{document}

\preprint{
\begin{minipage}[b]{1\linewidth}
\begin{flushright}
APCTP-PRE2015-014, IPMU15-0084\\
FTUV-15-06-25, IFIC-15-38\\
KIAS P15029
\end{flushright}
\end{minipage}
}

\title{Beyond the Dark matter effective field theory and \\ a simplified model approach at colliders}

\author{Seungwon Baek}
\email[]{sbaek1560@gmail.com}
\author{P. Ko}
\email[]{pko@kias.re.kr}
\affiliation{School of Physics, KIAS, Seoul 130-722, Korea}
\author{Myeonghun Park}
\email[]{parc@apctp.org}
\affiliation{ Asia Pacific Center for Theoretical Physics, San 31, Hyoja-dong, \\
  Nam-gu, Pohang 790-784, Republic of Korea}
  \affiliation{Department of Physics, Postech, Pohang 790-784, Korea}
\affiliation{Kavli IPMU (WPI), The University of Tokyo, Kashiwa, Chiba 277-8583, Japan}
\author{Wan-Il Park}
\email[]{Wanil.Park@uv.es}
\affiliation{Departament de Fisica Teorica and IFIC, Universitat de Valencia-CSIC, E-46100, Burjassot, Spain}
\author{Chaehyun Yu}
\email[]{chaehyun@gate.sinica.edu.tw}
\affiliation{Institute of Physics, Academia Sinica, Nangang, Taipei 11529, Taiwan}

\begin{abstract}

Direct detection of and LHC search for the singlet fermion dark matter (SFDM) model with Higgs portal interaction are considered in a renormalizable model where the full Standard 
Model (SM) gauge symmetry is imposed by introducing a singlet scalar messenger.  
In this model, direct detection is described by an effective operator 
$m_q \bar{q} q \bar{\chi} \chi$ as usual, but the full amplitude for monojet + \ET  involves
two intermediate scalar propagators, which cannot be seen within the effective field theory
(EFT) or in the simplified model without the full SM gauge symmetry.  
We derive the collider bounds from the ATLAS monojet + \ET as well as the CMS 
$t\bar{t}$ + \ET data, finding out that the bounds and the interpretation of the results 
are completely different from those obtained within the EFT or simplified models. 
It is pointed out that it is important to respect unitarity, renormalizability and local gauge invariance of the SM.
\end{abstract}

\maketitle

{\it 1. Introduction:}
Dark matter effective field theory (DM EFT) has been widely used for analyzing the data from DM (in)direct detections,  thermal relic density and collider searches for mono $X$ + \ET 
(with $X=g, \gamma, W,Z$ etc.)~\cite{monojet},   based on crossing symmetry which is an exact symmetry of quantum field theory.  Then very soon some limitation of DM EFT has 
been realized,  and  simplified DM models were introduced where mediators 
between the SM particles and DM are included explicitly~\cite{break,Buckley:2014fba}. 

In this letter, we point out that it is crucial to impose the full SM gauge symmetry, renormalizability and unitarity (including gauge anomaly cancellation) of the underlying 
DM models in order to study collider  signatures of DM 
and demonstrate  why the usual complementarity arguments break down for DM EFT 
or simplified DM models without the full SM gauge symmetry. Importance of the full SM 
gauge  invariance 
in the context of $u \bar{d} \rightarrow W^+$ + \ET was recently pointed out in 
Ref.~\cite{Bell:2015sza}. In this paper we consider the $q\bar{q} \rightarrow \chi\bar{\chi}$ which does not involve a massive weak gauge boson, and show that the full SM gauge 
symmetry is still important. 

As an explicit example, we will consider a singlet fermion DM model with 
Higgs portal interaction~\cite{Baek:2011aa,Baek:2012uj}, but we note that 
the arguments in this paper also work for vector DM with Higgs 
portal~\cite{Hambye:2008bq,Farzan:2012hh,Baek:2012se,Baek:2013dwa}.  
Then we repeat the analysis on the ATLAS monojet +  \ET signature~\cite{Aad:2015zva} 
and the CMS $t\bar{t}$ + \ET \cite{Khachatryan:2015nua} signatures  
and derive the bounds on the new physics scales, which are completely different from 
those obtained by ATLAS and CMS Collaborations \footnote{
See Ref.~\cite{Baek:2014jga} for the invisible branching ratios of the Higgs boson vs. 
direct detection cross sections within the EFTs and the UV completions of Higgs portal DM.}. 

Let us consider a scalar $\times$ scalar operator describing the direct detection of DM on nucleon,  assuming the DM is a Dirac fermion $\chi$ with some conserved quantum 
number stabilizing $\chi$: 
\begin{equation}
{\cal L}_{SS} \equiv \frac{1}{\Lambda^2} \bar{q} q \bar{\chi} \chi 
~~~{\rm or}~~~ \frac{m_q}{M_*^3} \bar{q} q \bar{\chi} \chi.
\label{eq:operator}
\end{equation}
Assuming the complementarity among direct detection, collider search and indirect detection (or thermal 
relic density),  the bound on the scale $\Lambda, (\textrm{or } M_*)$ of above operators has been studied extensively 
in literature~\cite{monojet}. 

However, the first operator in Eq.\,(\ref{eq:operator}) does not respect the full SM gauge symmetry and thus it may be
not suitable for studying phenomenology at high energy scale (say, at electroweak scale).    
Note that the SM quark bilinear part in the above operators can be written into 
$\overline{Q}_L H d_R ~~~{\rm or} ~~~\overline{Q}_L \widetilde{H} u_R$, 
if we impose the full SM gauge symmetry.  Here $Q_L \equiv ( u_L , d_L  )^T$. 
Thus one can write down the SM gauge symmetry invariant effective operator as a dimension 7 operator form;
\beq 
\mathcal L \ni \frac{y_d y_\chi}{\Lambda^3}\overline{Q}_L H d_R \,\bar \chi \chi+ \frac{y_u y_\chi'}{\Lambda^3}  \overline{Q}_L \widetilde{H} u_R \, \bar \chi \chi\, ,
\label{eq:operator2}
\eeq
so that the second operator in Eq.\,(\ref{eq:operator}) is from the above equation when the Standard Model $H$ gets the VEV.
Instead of above effective operator, one can also write down dimension 5 operator; 
\beq
\mathcal L \ni \frac{y_\chi}{\Lambda} H^+ H \,\bar \chi \chi \, ,
\eeq
since the singlet fermion $\chi$ cannot have renormalizable 
couplings to the SM Higgs boson. The interaction between quarks and dark matters is mediated by a Standard Molde Higgs in this case. 

The simplest way to write down a renormalizable operator that is invariant 
under the full SM gauge group is to introduce a real singlet scalar 
field $S$~~\cite{Baek:2011aa,Baek:2012se}
and induce an operator
$s \bar{\chi} \chi \times h \bar{q} q \rightarrow  \tfrac{1}{m_s^2} \bar{\chi} \chi \bar{q} q $ by integrating out the real scalar $s$.
However there is always a mixing between the SM Higgs $h$ and the real singlet scalar $s$, which results in two physical neutral scalars
$H_1$ and $H_2$ with the mixing angle $\alpha$. 
Therefore one should take into account the exchange of both $H_1$ and $H_2$
for DM direct detection scattering~\cite{Baek:2011aa}.
Note that there is a generic cancellation between two contributions from
two neutral scalars, which cannot be seen 
within EFT approach~\cite{Baek:2011aa,Baek:2012se}.

{\it 2. Renormalizable Model with the full SM gauge symmetry:}

The $s$-channel UV completion of the singlet fermion DM with Higgs portal 
has been constructed in Ref.~\cite{Baek:2011aa}:  
\begin{eqnarray} \label{eq:portal}
{\cal L}&=& \overline{\chi} ( i \slashed{\partial} - m_\chi - \lambda S ) \chi  
+ \frac{1}{2} \partial_\mu S \partial^\mu S  - \frac{1}{2} m_0^2 S^2
\\ \nonumber 
&-&   \lambda_{HS} H^\dagger H S^2 - \mu_{HS} S H^\dagger H - \mu_0^3 S
- \frac{\mu_S}{3 !} S^3 - \frac{\lambda_S}{4 !} S^4.
\end{eqnarray}
We note that the model defined by Eq.~(\ref{eq:portal}) is one possible UV completion
of the singlet fermion DM with effective interaction Eq.~(\ref{eq:operator})
\footnote{There is another type of UV completion with $t$-channel mediator, which is similar to 
SUSY models, and the LHC phenomenology of the model would be different from the
one considered in this paper.}.

Expanding both fields around their VEVs 
$(\langle H^0 \rangle = v_H, \langle S \rangle =v_s$), 
we can derive the Lagrangian in terms 
of $h$ and $s$. After diagonalization of the mass matrix, DM $\chi$ couples
with both $H_1$ and $H_2$.  

The interaction Lagrangian of $H_1$ and $H_2$ with the SM fields and DM $\chi$ is 
given by 
\begin{widetext}
\begin{equation} \label{eq:portal2}
{\cal L}_{\rm int} = - ( H_1 \cos\alpha + H_2 \sin\alpha ) \left[ \sum_f \frac{m_f}{v_H} 
 \bar{f} f  - \frac{2 m_W^2}{v_H} W_\mu^+ W^{-\mu} - \frac{m_Z^2}{ v_H} Z_\mu Z^\mu   
 \right] + \lambda ( H_1 \sin\alpha - H_2 \cos\alpha ) \bar{\chi}\chi \ ,
 \end{equation}
 \end{widetext}
following the convention of Ref.~\cite{Baek:2011aa}.  We identify the observed 125 GeV
scalar boson as $H_1$.  The mixing between $h$ and $s$ leads to the universal 
suppression of the Higgs signal strengths at the LHC, 
independent of production and decay channels ~\cite{Baek:2011aa}.

Let us start with the DM-nucleon scattering amplitude at parton level, 
$\chi (p) + q (k) \rightarrow \chi (p') + q (k')$, the parton level amplitude of which is given by 
\begin{widetext}
\begin{eqnarray}
{\cal M} & = & - \overline{u(p')} u(p) \overline{u(k')} u(k) ~\frac{m_q}{v_H} \lambda \sin\alpha \cos\alpha ~
\left[  \frac{1}{t - m_{H_1}^2 + i m_{H_1} \Gamma_{H_1}} - \frac{1}{t - m_{H_2}^2 + i m_{H_2} \Gamma_{H_2}}
\right] \label{eq:tchannel}   \\
&  \rightarrow &  \overline{u(p')} u(p) \overline{u(k')} u(k)  ~\frac{m_q}{2 v_H}  \lambda \sin 2\alpha 
\left[  \frac{1}{m_{H_1}^2}  - \frac{1}{m_{H_2}^2}    \right] 
\equiv \frac{m_q}{\Lambda_{dd}^3} \overline{u(p')} u(p) \overline{u(k')} u(k),
\label{eq:relation}
\end{eqnarray}
\end{widetext}
where $t \equiv (p' - p)^2$ is  the square of the 4-momentum transfer 
to the nucleon, and we took the limit
$t\rightarrow 0$ in the second line, which is a good approximation to the DM-nucleon scattering.
The scale of the dim-7 effective operator, $m_q\, \bar{q} q\, \overline{\chi}\chi$, describing the direct detection cross section for the DM-nucleon scattering is defined in terms of 
$\Lambda_{dd}$:
\begin{eqnarray}
\Lambda_{dd}^3 & \equiv & \frac{2 m_{H_1}^2 v_H}{ \lambda  \sin 2\alpha} \left(  1 - \frac{m_{H_1}^2}{m_{H_2}^2}  
\right)^{-1},  \\
\bar{\Lambda}_{dd}^3 & \equiv & \frac{2 m_{H_1}^2 v_H}{ \lambda  \sin 2\alpha},
\label{eq:BarlambdaDD}
\end{eqnarray}
where $\bar{\Lambda}_{dd}$ is derived from $\Lambda_{dd}$ in the limit 
$m_{H_2} \gg  m_{H_1}$.
It is important to notice that the amplitude~(\ref{eq:tchannel}) 
was derived from renormalizable and unitary 
Lagrangian with the full SM gauge symmetry, and thus can be a good starting point for addressing
the issue of validity of complementarity. 

The amplitude for the monojet with missing transverse energy($\met$) signature at hadron colliders is 
connected to the amplitude~(\ref{eq:tchannel}) by crossing symmetry  $s\leftrightarrow t$.  
Comparing with the corresponding  amplitude from the EFT approach, we have to include 
the following form factor:  
\begin{widetext}
\begin{equation}
\frac{1}{\Lambda^3_{dd}} \rightarrow \frac{1}{\bar{\Lambda}^3_{dd}} \left[ 
\frac{m_{H_1}^2}{ \hat{s} - m_{H_1}^2 + i m_{H_1} \Gamma_{H_1}}  - 
\frac{m_{H_1}^2}{ \hat{s} - m_{H_2}^2 + i m_{H_2} \Gamma_{H_2}} \right]  
\equiv  \frac{1}{\Lambda^3_{col} ( \hat{s} )}  ,
\label{eq:collisionS}
\end{equation}
\end{widetext}
where $\hat{s} \equiv M_{\chi\chi}^2$ is  the square of the invariant mass  of the DM pair.
Note that $\hat s \geq 4 m_\chi^2$ in the physical 
region for DM pair creation,  and that there is no single constant scale  
$\Lambda_{col}$  for an effective operator that characterizes the 
$q\bar{q} \rightarrow \chi \bar{\chi}$, since $\hat{s}$ varies in the range of 
$4 m_\chi^2 \leq \hat{s} \leq s$ with $\sqrt{s}$ being the center-of-mass (CM) energy 
of the collider.   
Also note that we have to include two scalar propagators with opposite sign in order to 
respect the full SM gauge symmetry  and renormalizability. This is in sharp contrast with 
other  previous studies where only a single propagator is introduced to replace 
$1/\Lambda^2$.    
The two propagators interfere destructively for very high $\hat{s}$
or small $t$ (direct detection), but for $m_{H_1}^2 < \hat{s} <m_{H_2}^2$,
they interfere constructively.
The $1/s$ suppressions from the $s$-channel resonance 
propagators  make the amplitude unitary, in compliance with renormalizable and unitary QFT.

If one can fix $\hat{s}$ and $m_{H_2}^2 \gg \hat{s}$, we can ignore  the  2nd propagator. 
But at hadron colliders, $\hat{s}$ is not fixed, 
except for the kinematic condition $4 m_\chi^2 \leq \hat{s} \leq s$ 
(with $s = 14$TeV for example at the LHC@14TeV).   
Therefore we cannot say clearly when we can ignore $\hat{s}$ compared with 
$m_{H_2}^2$ at hadron colliders, unless $m_{H_2}^2 > s$ (not $\hat{s}$).  

 \begin{figure*}[th!]
 \centering
\includegraphics[width=0.35\textwidth]{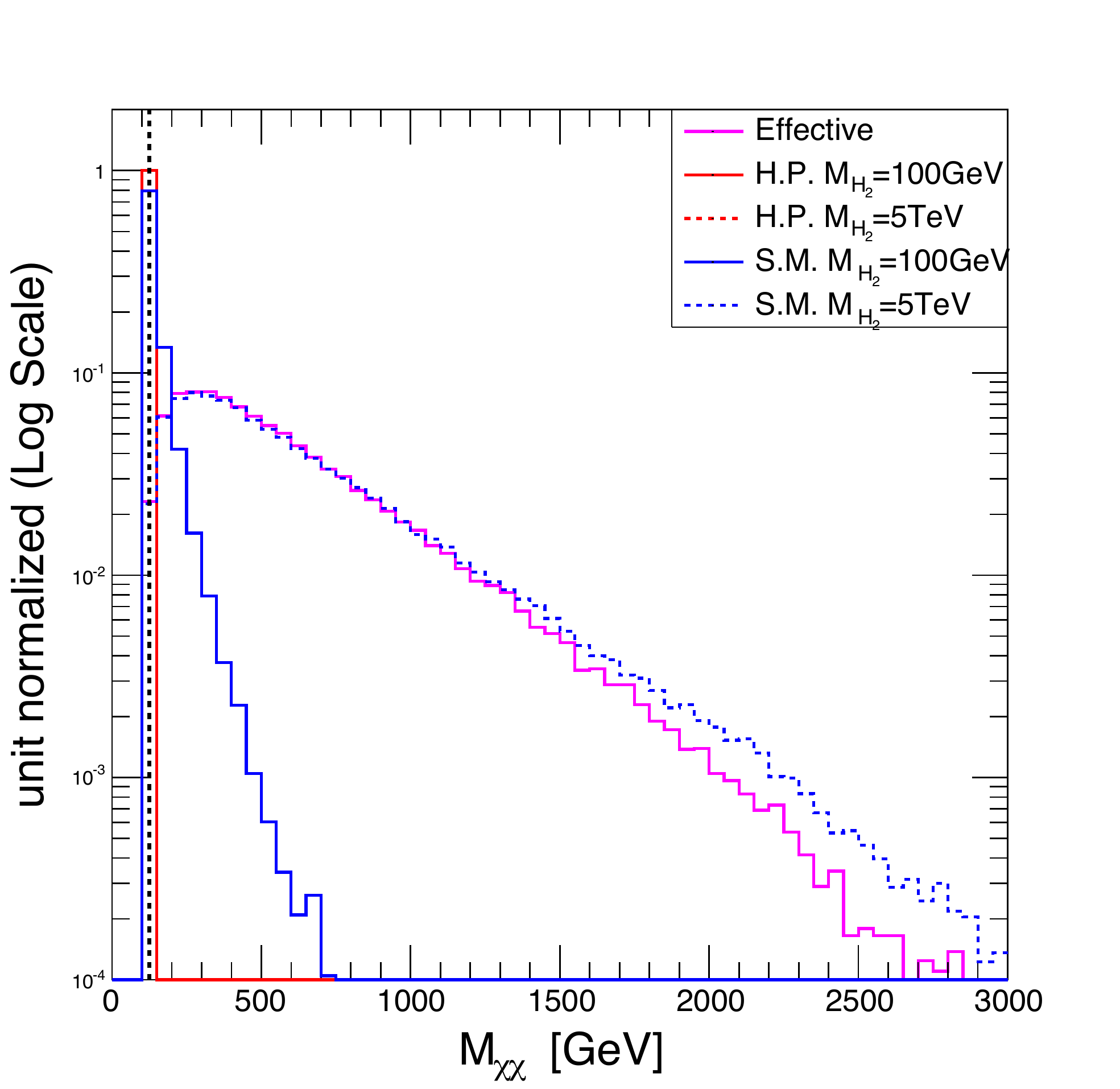}
\includegraphics[width=0.35\textwidth]{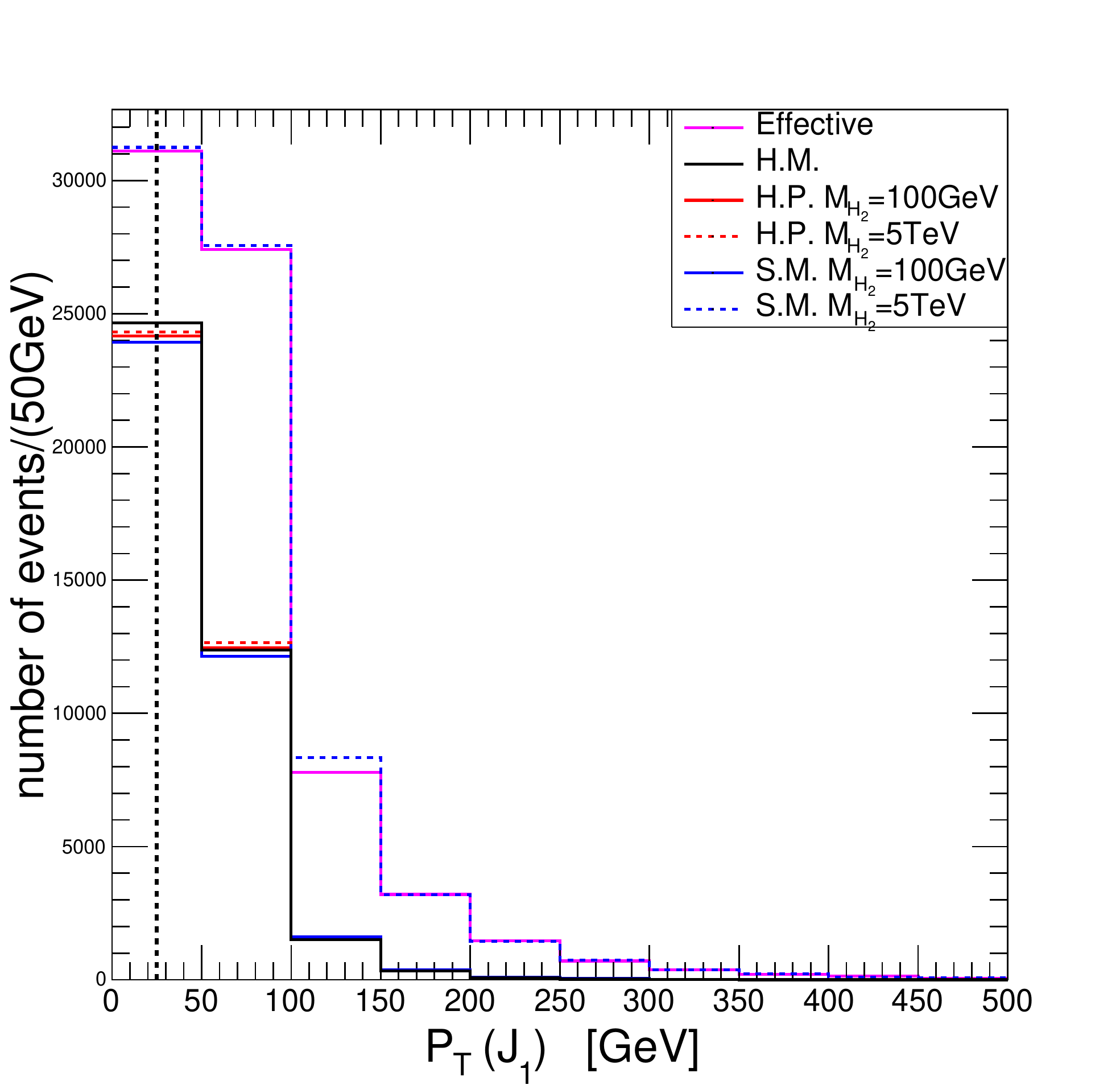}\\
\includegraphics[width=0.35\textwidth]{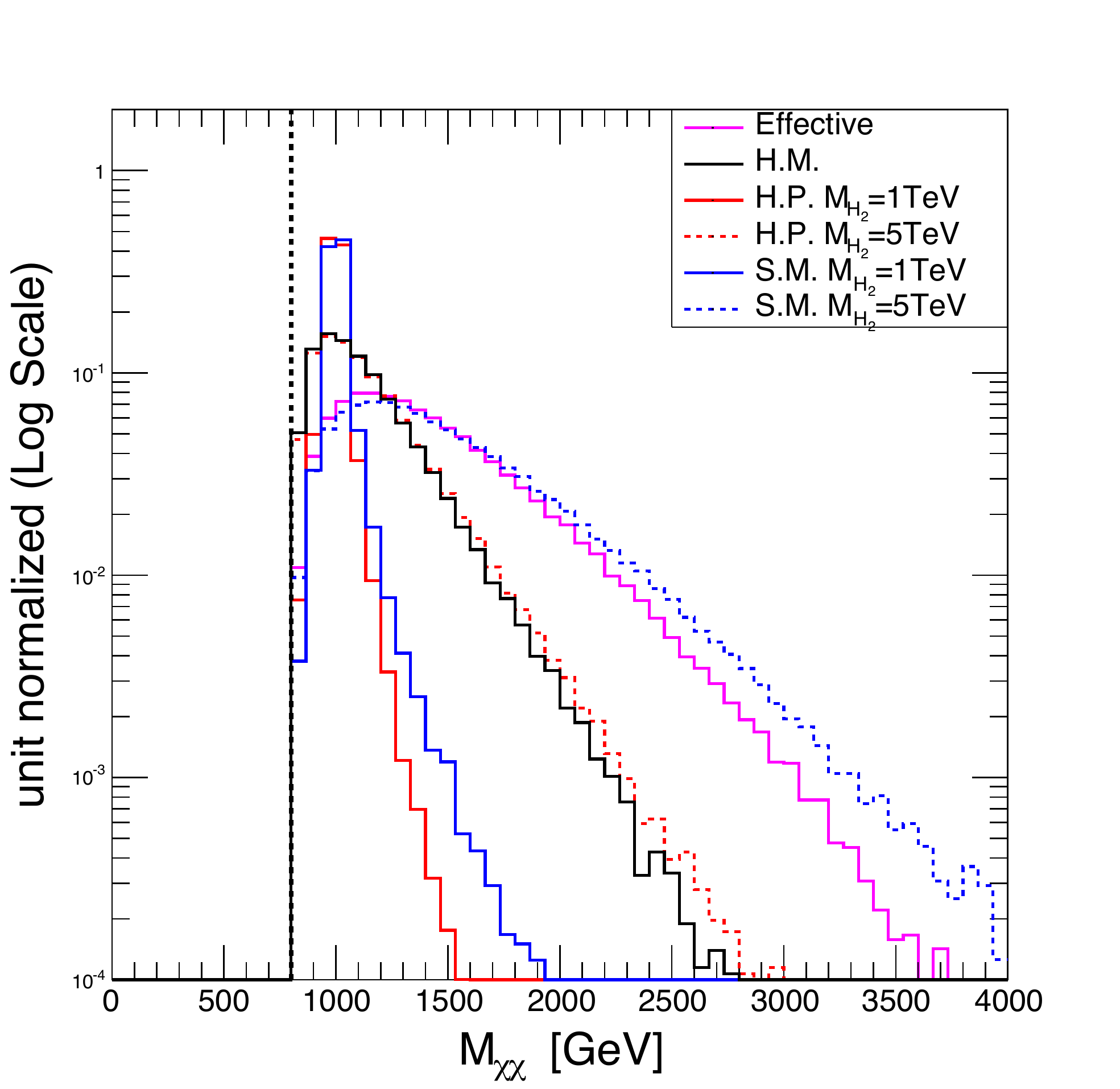}
\includegraphics[width=0.35\textwidth]{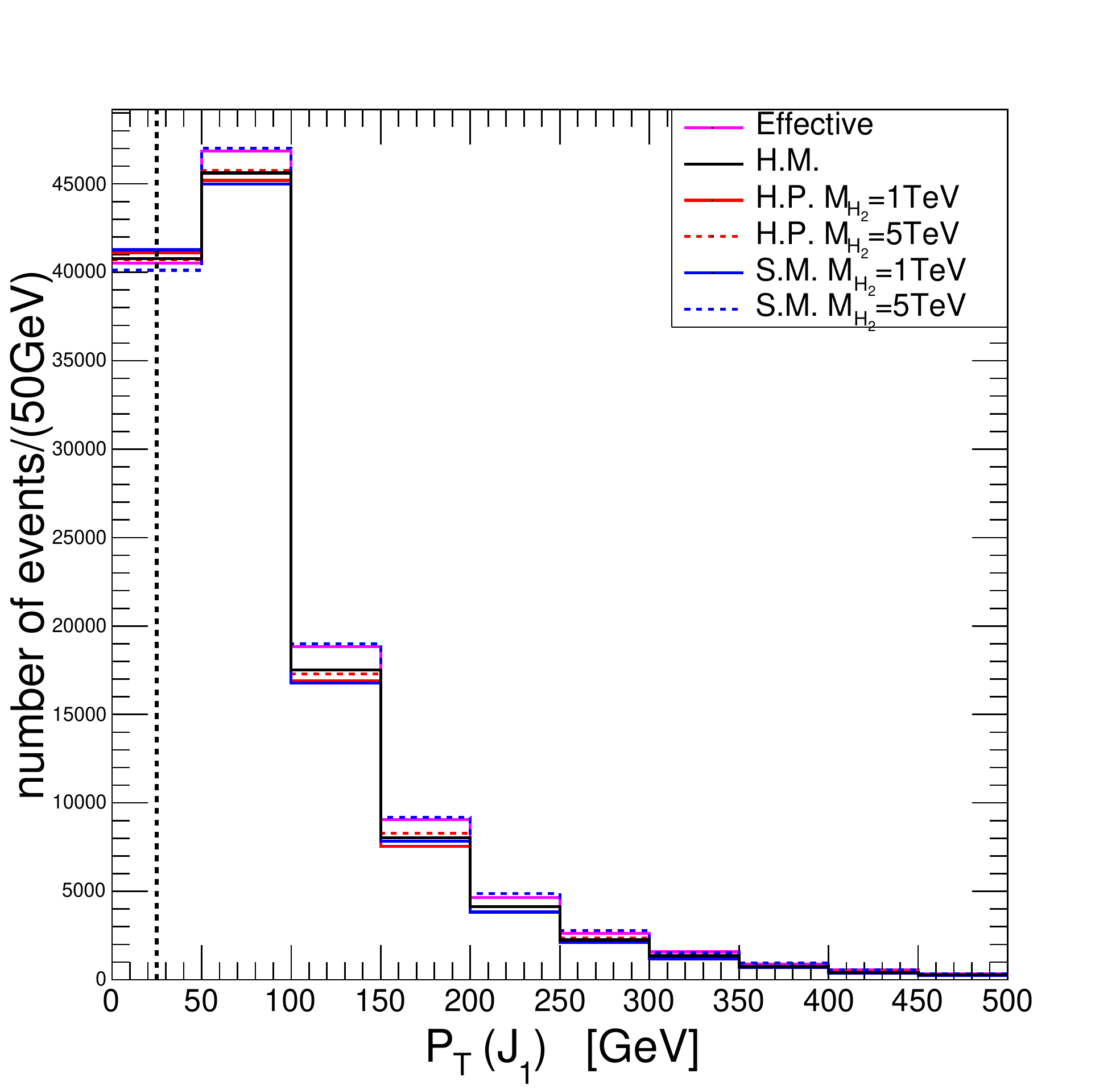}
\caption{\label{fig:monoKin} 
$M_{\chi\chi}$ and $P_T$ distribution of the hardest jet 
in the reconstruction level (after the detector simulation)
for $m_\chi=50$ GeV (upper panels)
and $m_\chi=400$ GeV (lower panels)
in the ATLAS 8 TeV monojet+$\met$ search.
}
\end{figure*}

{\it 3. Collider Studies:}  
There are two important factors in the search for new physics at colliders: a total cross section and the shape of differential cross sections with respect to various analysis ``cut" variables.  A mixing angle $\alpha$ between two scalars is related only to a total cross section, not to the shape of differential cross section.
The shape of differential cross sections and efficiencies from various analysis cuts are related to the nontrivial propagators coming 
from two mediators $(H_1, H_2)$. Thus we can single out the effect of a mixing angle from collider analyses when we try to understand whether we can recast 
results of various analyses based on the effective operator and a simplified model to our model here, the Higgs portal case through the following set up:

\begin{itemize}
\item{EFT} : Effective operator 
$\mathcal{L}_{int}=\frac{m_q}{M_*^3}\bar{q}q\bar{\chi}\chi$ defined in Eq. (1)
\item{S.M.}: Simplified model with a scalar mediator $S$ \cite{Buckley:2014fba}, \\[1mm]
$\mathcal{L}_{int}=\left(\frac{m_q}{v_H}\sin{\alpha}\right) s \bar{q}q-\lambda  s \bar{\chi}\chi \cos{\alpha}$
\item{H.M.}: A Higgs boson as a mediator,  \\[1mm]
$\mathcal{L}_{int}=-\left(\frac{m_q}{v_H}\cos{\alpha}\right) h \bar{q}q-\lambda  h \bar{\chi}\chi \sin{\alpha}$
\item{H.P.}: Higgs portal model defined in Eq.~(\ref{eq:portal}) or (\ref{eq:portal2}). 
\end{itemize}
In the S.M. and H.M. cases, we can regard $\alpha$ as a suppression factor in interactions 
while in the H.P. case, it is a mixing angle between $h$ and $s$.
Note that the SM gauge symmetry is not fully respected within EFT, S.M. and H.M. cases. 

The kinematics of a signature, i.e., $P_T$ of an initial state radiation (ISR) jet and the size of $\met$,  depend on the scale of a hard interaction, 
which is proportional to the invariant mass of a dark matter pair, $M_{\chi\chi}$.  
With following LHC studies, we show that there are relations among EFT, S.M., H.M., and H.P:
\begin{eqnarray}
\textrm{H.P.} \underset{m_{H_2}^2 \gg \hat{s}}{\longrightarrow} \textrm{H.M.},
\label{eq:HPtoHM} \\
\textrm{S.M.} \underset{m_S^2 \gg \hat{s}}{\longrightarrow} \textrm{EFT}, 
\label{eq:SMtoEFF} \\
\textrm{H.M.} \neq \textrm{EFT}\, .
\label{eq:HMtoEFF}
\end{eqnarray}
In H.P., the limit $m_{H_2}^2 \gg \hat{s}$ can be achieved, for example, by taking $v_S$ (the VEV of $S$ in \eq{eq:portal}) large while keeping dimensionless couplings perturbative.
The mixing angle in this case is approximated to \cite{Baek:2012uj}
\beq \label{alpha}
\tan 2 \alpha \simeq \frac{2 v_H \l( \mu_{H_S} + \lambda_{HS} v_S \r)}{2 \lambda_S v_S^2}\, .
\eeq
The perturbativity of effective couplings obtained after integrating out the heavy scalar particle ($H_2$) requires $\mu_{HS} + \lambda_{HS} v_S \lesssim m_{H_2}$, constraining the mixing angle to be upper-bounded as 
\beq \label{alpha-bnd}
\alpha \lesssim 2 \sqrt{\frac{\pi}{3}} \frac{v_H}{m_{H_2}} .
\eeq
Hence, as $H_2$ becomes heavier, impacts of H.P. at collider experiments becomes more elusive.  In any case, for $m_{H_2}^2 \gg \hat{s}$, the effect of the heavy scalar 
propagator can be ignored in relevant diagrams for collider searches. 
Then, it is clear that H.P. reduces to H.M. with the angle $\alpha$ given by \eq{alpha}, 
and this is what \eq{eq:HPtoHM} means.  On the other hand, it should be clear that, 
S.M. is reduced to EFT for $m_S^2 \gg \hat{s}$, as stated in Eq.\,(\ref{eq:SMtoEFF}),
since there is only one scalar mediator which can be very heavy in S.M. 
\footnote{However, in this case, $\Lambda_{dd}$ and $m_{H_1}$ should be replaced by 
$\overline{\Lambda}_{dd}$ and $m_s$.}.
Also, it should be clear that, since the mass of SM-like Higgs is fixed, H.M. cannot 
be reduced to EFT for $m_h^2 \lesssim \hat{s}$, as stated in Eq.(\ref{eq:HMtoEFF}).

Thus, an effective operator approach cannot capture the feature of an actual dark matter model, as shown here in the context of the Higgs portal singlet fermion DM as an example. 
We illustrate our point with the ATLAS monojet and the CMS 
$t\bar{t}+\met$ searches \cite{Aad:2015zva, Khachatryan:2015nua}.
We simulate LHC  8TeV events with Madgraph, Pythia 6 and Delphes simulations\,\cite{montecarlos,Sjostrand:2006za, deFavereau:2013fsa}. For the ATLAS monojet search, we use MLM-scheme\,\cite{MLM} to match one jet with a matching scale around
$\left(\frac{1}{6}\sim\frac{1}{3}\right)\times\sqrt{\met^2+4m^2_\chi}$. For a width of a scalar mediator, 
we take $\Gamma_S = m_S/(8\pi)$ \cite{Aad:2015zva}.

{\it 3.1 Monojet + \ET signatures:} 
We adopt the selection cuts in ATLAS monojet search \cite{Aad:2015zva}.
Depending on $\met$, ATLAS has 9 signal regions, from $\met > 150\GeV$ to 
$\met>700\GeV$. 
The hardness of ISR is proportional to the energy scale of a dark matter pair $M_{\chi\chi}$, 
which is depending on propagator(s) of mediators. 
To illustrate this feature more clearly, 
we show distributions of $M_{\chi\chi}$ and $P_T$ of a leading jet in Fig.~\ref{fig:monoKin} with study points
of $(m_S, m_{H_2}) \in \{100\GeV, 5\TeV\}$ when a dark matter mass is $50\GeV$ 
and $(m_S, m_{H_2}) \in \{1\TeV, 5\TeV\}$ for $m_\chi= 400\GeV$.
We selected events with no isolated leptons and at least with hard jet of 
$P_T>30\GeV$ at the detector level. 
If the poles of propagators (the mass of a mediator) are within the reach of $M_{\chi\chi}$, 
the kinematics are fixed at the mass scale of mediators.
For an example, when $m_\chi=50\GeV$, kinematics are localized at the mass of Higgs in the H.M. and H.P. cases. 
In the S.M. case with $m_S=100\GeV$, due to the finite size of $\Gamma_S$, $M_{\chi\chi}$ distribution becomes wide compared to H.M. and H.P. cases. 
Thus ISR in S.M. case become slightly larger compared to H.M. and H.P. cases 
so that the efficiency of analyses  becomes larger correspondingly. 
When a mediator mass $m_S=5\TeV$ in S.M. case, the pole in a mediator's propagator 
is far away at $\sqrt{s}=8$ TeV so that the kinematics become similar to the EFT case 
where there is no pole structure~\cite{Aad:2015zva}.
Similarly, when $m_\chi=400$ GeV, In H.P. with $m_{H_2}=1\TeV$, the propagator from 
the Higgs does not contribute to the kinematics since it is located below to the threshold 
of dark matter pair production. Thus kinematics is same as S.M. with $m_S=1$ TeV.  
But when $m_{H_2}$ is large  enough compared to the threshold of dark matter pair 
production $(m_{H_2} \sim 5\TeV)$, 
the major effect of propagators is from the Higgs propagator. 
This feature also appears in the H.M. case. 

It is evident that for $m_\chi=50$ GeV, 
the EFT and the S.M. with $m_S=5$ TeV show similar
distributions while H.M and H.P also do.
For $m_\chi=400$ GeV, the same features are shown
between the EFT and S.M. with $m_S=5$ TeV,
between H.M. and H.P. with $m_{H_2}=5$ TeV,
and between S.M. with $m_S=1$ TeV and H.P. with $m_{H_2}=1$ TeV,
respectively.

\begin{figure}[th]
\includegraphics[width=0.43\textwidth]{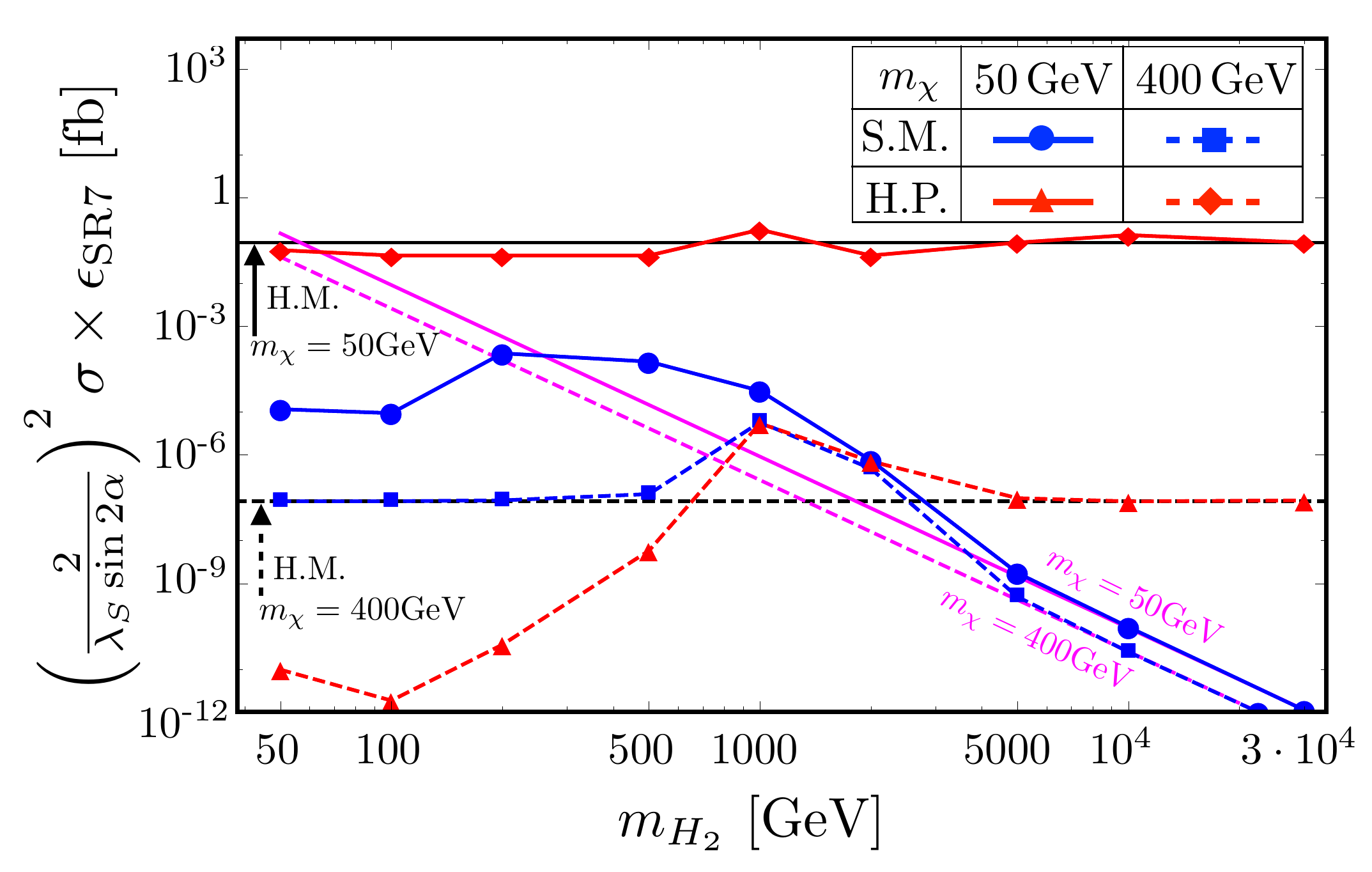}
\caption{
Rescaled cross sections for the monojet+$\met$ 
in the signal region SR7 ($\met> 500\GeV$) at ATLAS~\cite{Aad:2015zva}.
Each line corresponds to the EFT approach (magenta), S.M. (blue),
H.M. (black), and H.P. (red), respectively. 
The solid and dashed lines correspond to $m_\chi=50$ GeV and $400$ GeV 
in each model, respectively.
}
\label{fig:monoCross}
\end{figure}

Final search results will also depend on the production cross section which depends on propagators of mediators.
In Fig.~\ref{fig:monoCross}, we illustrate the cross sections 
rescaled by the dimensionless factor $\left(2/\lambda_S \sin 2\alpha \right)^2$
and the efficiency $\epsilon_{\textrm{SR7}}$ in the signal region SR7
($\met> 500\GeV$) at ATLAS~\cite{Aad:2015zva}.
The rescaled cross sections are apparently independent of
the mixing angle $\alpha$.
The figure clearly shows that the Higgs portal model cannot be described by
either the EFT or the S.M at all.
Also in the limit that $m_{H_2} (m_S)$ is much larger than the typical scale
in the process, the S.M approaches the EFT, whereas the H.P. does the H.M., respectively. 

\begin{figure}[th]
\includegraphics[width=0.43\textwidth]{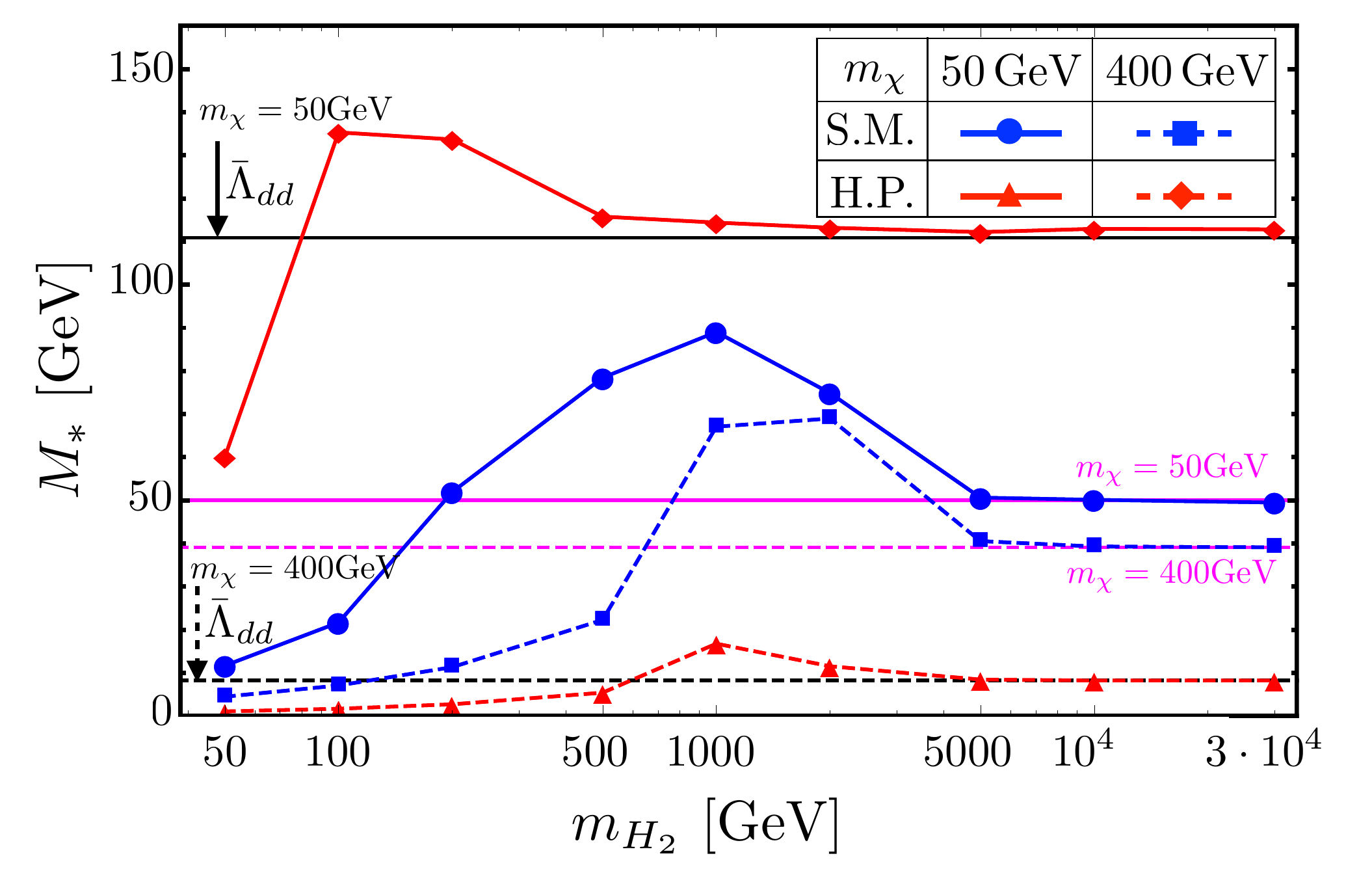}
\includegraphics[width=0.43\textwidth]{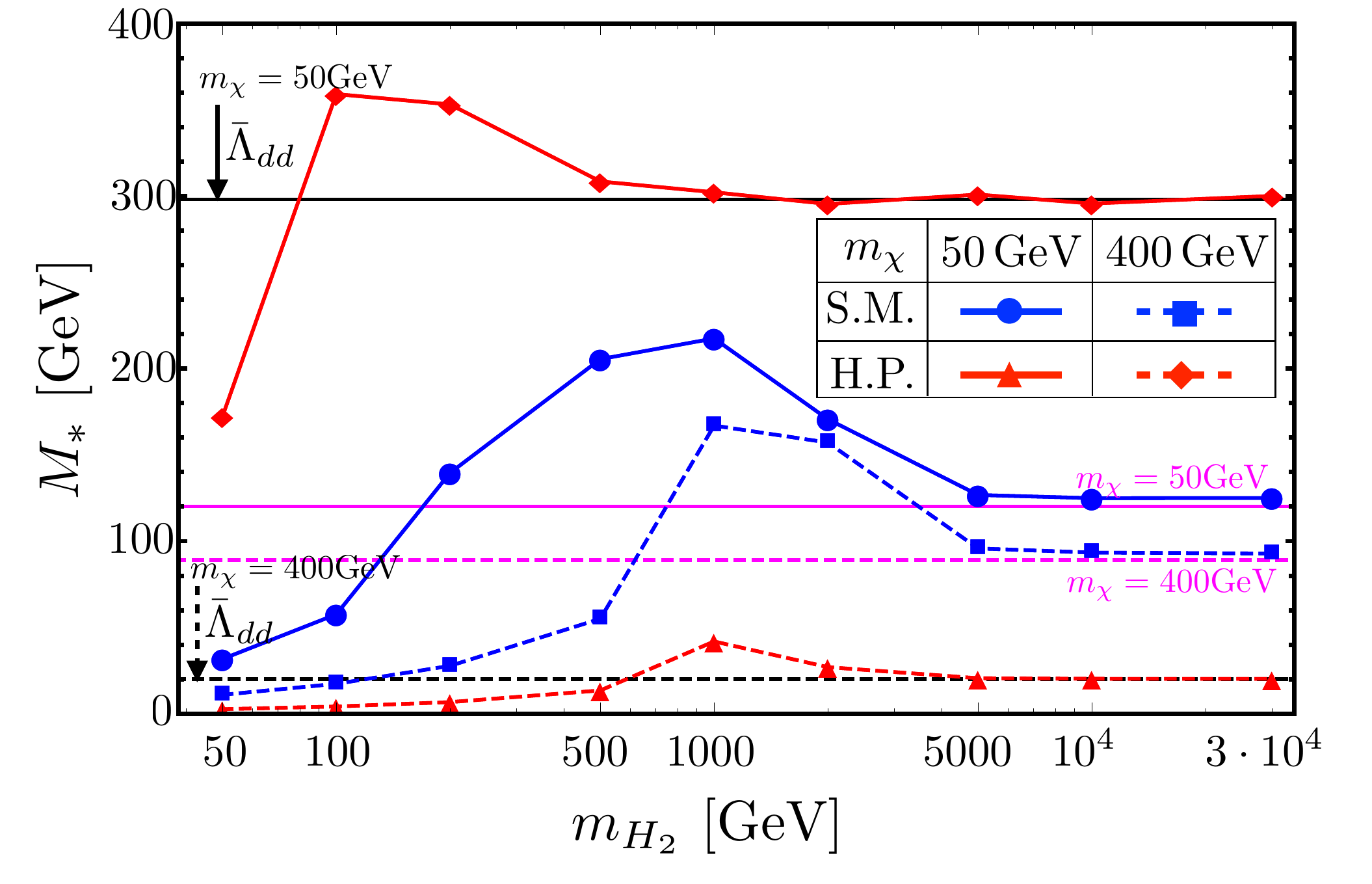}
\caption{
The experimental bounds on $M_*$ at $90\%$ C.L. as a function of $m_{H_2}$ ($m_S$ in S.M.\,case)
in the monojet+$\met$ search (upper) and $t\bar t+\met$ search (lower).
Each line corresponds to the EFT approach (magenta), S.M. (blue),
H.M. (black), and H.P. (red), respectively. 
The bound of S.M., H.M., and H.P., are expressed in terms of the effective mass $M_*$ through the Eq.(\ref{eq:comp})-(\ref{eq:comp2}).
The solid and dashed lines correspond to $m_\chi=50$ GeV and $400$ GeV
in each model, respectively. }
\label{fig:simp}
\end{figure}

{\it 3.2 $t\bar{t}$ + \ET signatures:}
A (effective) scalar operator in Eq.~(\ref{eq:operator})  from the Higgs portal case is proportional to the 
mass of quarks. Thus dark matter creations with top quark pair will have better sensitivities compared to the usual monojet search\,\cite{Lin:2013sca,Haisch:2015ioa}.  
Following the analysis of CMS $t\bar{t}+\met$ search~\cite{Khachatryan:2015nua},
we find similar features in the monojet search in the previous section.
The detail of this analysis will be presented in 
the future publication~\cite{future}, 
but we will show the resulting bound on $M_*$ in Fig.~\ref{fig:simp} 
(the lower pannel) in the following subsection.

{\it 3.3 Relation between a mediator and an effective operator approach:}
By direct comparison between scattering matrix elements from an effective operator 
and from a simple scalar mediator, 
we can have a similar relation to Eq.~(\ref{eq:BarlambdaDD}) 

\beq
M_*^3= \left(\frac{2 v_H}{\lambda \sin{2\alpha}}\right) m_S^2.
\label{eq:comp}
\eeq
With this relation, the ATLAS collaboration showed that the validity of the effective 
operator when $m_S>5\TeV$ \cite{Aad:2015zva}. 
However as shown in Eq.~(\ref{eq:SMtoEFF}), this validity holds only for the S.M which
does not respect the full SM gauge symmetry,  while the H.P. with the full SM gauge 
symmetry does not approach the EFT result.

In Fig.~\ref{fig:simp}, we show that the experimental $90\%$ C.L.\,limits on the suppression scale $M_*$ as a function of a mediator mass $m_{H_2}$ ($m_S$ in the S.M.\,case) 
at the LHC by using the results in the monojet+$\met$ search (upper) at ATLAS \cite{Aad:2015zva}
and in the $t\bar t+\met$ search (lower) at CMS \cite{Khachatryan:2015nua}.
For the translation from the limit on the mass of a mediator in a specific model to a limit on the $M_*$ in the effective operator, 
we use a direct comparison between parameters in a model and an suppression scale $M_*$ in the limit where a collision energy becomes negligible compared to the mediator's mass. For S.M.\,case we use the following relation 
\beq
\frac{m_q}{M_*^3} = \frac{m_q \lambda \sin\alpha \cos\alpha}{v_H} \frac{1}{m_S^2} 
\eeq
so that a limit on $M_*$ can be obtained through a translation 
\beq
\left[ \left(\frac{1}{M_*^3}\right)^2 \left(\frac{\lambda \sin2\alpha}{2 v_H m_S^2}\right)^{-2}
 \sigma_{\textrm{(S.M.)}} \right]\times  \epsilon_{\textrm{(S.M.)}}= \frac{\textrm{N}_{obs}}{ \mathcal L}\, .
  \eeq
For a H.P.\,case, we use  
 \beq
\frac{m_q}{M_*^3} = \frac{m_q \lambda \sin\alpha \cos\alpha}{v_H} \left(\frac{1}{m_{H_1}^2}-\frac{1}{m_{H_2}^2}\right) = \frac{m_q}{\Lambda_{dd}^3} \, ,
  \eeq
to take care of a finite but nonzero term from the light mediator ($H_1$) in the limit of a heavy $H_2$. The bound on $M_*$ will be from 
 \beq
 \left[ \left(\frac{1}{M_*^3}\right)^2 \left(\frac{1}{\Lambda_{dd}^3}\right)^{-2} \sigma_{\textrm{(H.P.)}}\right] \times\epsilon_{\textrm{(H.P.)}} = \frac{\textrm{N}_{obs}}{ \mathcal L}\, , 
\label{eq:comp2}
\eeq
where $\mathcal L$ is the beam luminosity, $\epsilon$ is the total efficiency including the signal acceptance and $N_{obs}$ is the upper limit for the number of signal events reported by the LHC experimental collaborations. With above translation procedure, the mixing angle $\alpha$ dependency of the exclusion limit for H.P. from Eq.\,(\ref{alpha-bnd}) does not appear in the translated exclusion limit on $M_*$ through a cancelation by $\Lambda_{dd}$.

It is apparent that as the mediator mass $m_{H_2}$ becomes large, 
LHC analyses on H.P. becomes similar to H.M. as in Eq.\,(\ref{eq:HPtoHM}).
This feature clearly shows that the Higgs portal model cannot be described
by contact interaction even in the limit that the mass of the extra scalar boson
become large enough. The main reason is that even in that case,
there exists
the SM Higgs boson exchange arising from the mixing between two scalar bosons,
which cannot be described by contact interaction at the LHC.
It is also worthwhile to mention that as the $m_{H_2}(=m_S)$ in the S.M.
becomes heavy enough, the bound in the S.M. approaches that in the EFT approach.

Finally, we would like to note that the arguments in this paper are not to be applied directly 
to other UV completions.  
However, we emphasize that if the origin of the quark mass 
in Eq.~(\ref{eq:operator}) is the electroweak symmetry breaking, and the corresponding 
interaction of DM with a Higgs boson must also be taken into account in the energy scale of the LHC. 

{\it 4. Conclusion:}
In this letter, we discussed the DM search at LHC within a renormalizable theory for singlet 
Dirac fermion DM with Higgs portal  and the full SM gauge symmetry, and demonstrated 
how and why the EFT approach or DM simplified models without the full SM gauge 
symmetry can break down for collider searches,  and why the complementarity based on 
crossing symmetry of quantum field theory can be misleading. 
We also reanalyzed the ATLAS monojet data and the CMS $t\bar{t}$ + \ET and derived 
new bounds on the new physics scale $M_*$, which are completely different from the 
bounds presented by   ATLAS and CMS Collaboration \footnote{See Ref.
~\cite{Freitas:2015hsa} for  the production of the 2nd scalar at the LHC, 
which is complementary to the \ET signatures we considered in this paper.}.  
The origin of this difference is the full SM gauge symmetry imposed on the UV completions, 
which is very important for DM model building and phenomenology therein 
\footnote{Unitarity is also important, as noticed in Ref.~\cite{Baek:2012se} in the context 
of massive VDM with Higgs portal.}.

\section{Note Added}
Preliminary results of this paper have been reported at a number of workshops~\cite{link}  
by one of the authors (PK),  emphasizing the roles of SM gauge symmetry and the form 
factor with two scalar propagators, Eq.\,(\ref{eq:collisionS}). 
While we are finishing this paper, there appeared a paper where 
these points were discussed to some extent \cite{Abdallah:2015ter}.  

\section{Acknowledgments}
We are grateful to R. Allahverdi, H. Baer, O. Buchmueller, J. B. Dent, B. Dutta, Chengcheng Han,
Sunghoon Jung, M. Kakizaki, T. Kamon,  A. Kotwal, Jungil  Lee, S. Matsumoto,  
T. J. Weiler  and Hai-Bo Yu for discussions on the topics in this paper and 
the related issues. 

This work is supported in part by National Research Foundation of Korea (NRF) 
Research Grant 
NRF-2015R1A2A1A05001869 (SB, PK), and by the NRF grant funded by the Korea government (MSIP) 
(No. 2009-0083526) through Korea Neutrino Research Center at Seoul National University (PK). 
MP is supported by the Korea Ministry of Science, ICT and Future Planning, Gyeongsangbuk-Do and 
Pohang City for Independent Junior Research Groups at the Asia Pacific Center for Theoretical Physics.
MP is also supported by World Premier International Research Center Initiative (WPI Initiative), MEXT, Japan.
WIP acknowledges support from the MEC and FEDER (EC) Grants FPA2014-54459 and the Generalitat 
Valenciana under grant PROMETEOII/2013/017.
The work of CY was supported by the Ministry of Science and Technology (MOST)
of Taiwan under grant number 101-2112-M-001-005-MY3.

\end{document}